\documentclass[11pt,a4paper]{article}
\usepackage{txfonts}
\usepackage{graphicx}
\usepackage{amssymb}

\title{All solutions of the $n = 5$ Lane--Emden equation}
\author{Patryk Mach\\
{}\\
\textit{M. Smoluchowski Institute of Physics, Jagiellonian University,}\\
\textit{Reymonta 4, 30-059 Krak\'{o}w, Poland}}
\date{}

\begin{document}
\maketitle

\begin{abstract}
All real solutions of the Lane--Emden equation for $n = 5$ are obtained in terms of Jacobian and Weierstrass elliptic functions. A new family of solutions is found. It is expressed by remarkably simple formulae involving Jacobian elliptic functions only. The general properties and discrete scaling symmetries of these new solutions are discussed. We also comment on their possible applications.
\end{abstract}

\section{Introduction}

The Lane--Emden equation
\begin{equation}
\label{len}
\Delta \theta + \theta^n = \frac{1}{\xi^2} \frac{d}{d \xi} \left( \xi^2 \frac{d \theta}{d \xi} \right) + \theta^n = 0
\end{equation}
is a classical equation of mathematical physics. Introduced in 1870 by Lane~\cite{lane}, and studied in 1907 by Emden~\cite{emden}, it was originally used to describe the mass density distribution inside a (spherical) polytropic star in hydrostatic equilibrium (in that case, the Lane--Emden equation follows directly from the Poisson equation for the gravitational potential and the assumption of hydrostatics). Shortly after Emden's work was published, Plummer~\cite{plummer} proposed a model of the distribution of stars in globular clusters that was based on solutions of Eq.~(\ref{len}). Since that time, the Lane--Emden equation and its generalisations were applied in many different branches of physics, ranging from astrophysics to kinetic theory and quantum mechanics (see, e.g., \cite{goenner} and references therein).

\tolerance=1000
There are 2 well known solutions of the $n=5$ Lane--Emden equation, i.e., 
\begin{equation}
\label{le}
\frac{1}{\xi^2} \frac{d}{d \xi} \left( \xi^2 \frac{d \theta}{d \xi} \right) + \theta^5 = 0.
\end{equation}
The first of them,
\begin{equation}
\label{schuster}
\theta(\xi) = \pm \frac{1}{\sqrt{1 + \xi^2/3}},
\end{equation}
is due to Schuster and Emden~\cite{emden}. The second,
\begin{equation}
\label{fixed}
\theta(\xi) = \pm \frac{1}{\sqrt{2 \xi}},
\end{equation}
which is singular at $\xi = 0$, belongs to a larger class of solutions valid for all Lane--Emden equations with $n > 3$, that is
\[ \theta(\xi) = \left( \frac{2 (n - 3)}{(n - 1)^2 \xi^2} \right)^{1/(n-1)}. \]
Because the Lane--Emden equation (\ref{le}) is invariant under the transformation $S\colon \theta(\xi) \to \theta(\xi/\lambda)/\sqrt{\lambda}$, $\lambda \in \mathbb R_+$, Schuster's integral (\ref{schuster}) yields, in fact, a whole one-parameter family of solutions. Solution (\ref{fixed}) is a fixed point of $S$.

It was believed for many years that finding other solutions to Eq.~(\ref{le}) `is complicated and involves elliptic integrals' (Chandrasekhar~\cite{chandrasekhar}). Then, in 1962 Srivastava~\cite{srivastava} found another solution that can be written in a compact form, namely
\begin{equation}
\label{sri}
\theta(\xi) = \pm \frac{\sin \left(\ln \sqrt{\xi} \right)}{\sqrt{3 \xi  - 2 \xi \sin^2 \left( \ln \sqrt{\xi} \right)}},
\end{equation}
where again the scaling transformation gives a whole family of solutions. This solution is also singular at $\xi = 0$, but it can be used in composite stellar models~\cite{murphy83}. It was rediscovered in 1977 by Sharma~\cite{sharma}, who searched for non-elliptic solutions of (\ref{le}).

About a decade ago Goenner and Havas~\cite{goenner} found the following expression for another class of solutions
\begin{equation}
\label{gh}
\theta(\xi) = \frac{c_1}{\sqrt{-\xi/3 + \xi \wp \left( \ln (B \xi)/2; 4/3, -8/27 + 16 c^4_1 / 3 \right)}},
\end{equation}
where $\wp$ denotes the Weierstrass elliptic function, and $B$ and $c_1$ are integration constants (cf.~Eq.~(30) of~\cite{goenner}). This simple result was somehow overlooked by researchers---see, e.g., a recent book by Horedt~\cite{horedt}, where the aforementioned opinion of Chandrasekhar is repeated.

In this paper we derive all remaining real solutions of Eq.~(\ref{le}) (new solutions are given by Eq.~(\ref{case1}) in Sec.~2.1). We also show that for a certain range of integration constants, solution (\ref{gh}) can be obtained in terms of Jacobian elliptic functions. Because of many different physical applications of Lane--Emden equations, we will not discuss any particular boundary conditions here.

\section{Solutions}

Equation (\ref{le}) can be transformed to an autonomous form by the following substitution: $\theta = z/\sqrt{2 \xi}$ and $t = - \ln \xi$. The resulting equation is
\[ \frac{d^2 z}{dt^2} = \frac{1}{4} z \left( 1 - z^4 \right), \]
and its standard integration yields
\begin{equation}
\label{dyn}
\left( \frac{dz}{dt} \right)^2 = \frac{1}{12} \left( -z^6 + 3z^2 + C \right),
\end{equation}
where $C$ denotes an integration constant (see e.g.~\cite{chandrasekhar}).

Further analysis depends on the factorisation of the polynomial $w(z) = -z^6 +3z^2 + C$ appearing in Eq.~(\ref{dyn}).
\begin{enumerate}
\item For $C < -2$ the polynomial $w(z)$ is negative. Thus, there are no real solutions of Eq.~(\ref{dyn}).
\item For $C = -2$ the polynomial $w(z) = -(z-1)^2 (z+1)^2 (z^2 + 2)$ is nonpositive, and it has two zeros at $z = \pm 1$. Setting $z \equiv \pm 1$ yields the singular solution (\ref{fixed}).
\item For $C \in (-2,0)$ the polynomial $w(z)$ has 4 real roots that can be found using Cardano's formulae. Denoting
\begin{eqnarray}
a & = & 2 \sin \left( \frac{1}{3} \arcsin \left(\frac{|C|}{2} \right) \right), \;\;\; b = 2 \cos \left( \frac{1}{3} \arccos \left( - \frac{|C|}{2} \right) \ \right), \nonumber \\
c & = & 2 \cos\left( \frac{1}{3} \arccos \left( \frac{|C|}{2} \right) \right),
\label{cardano}
\end{eqnarray}
we obtain $w(z) = (z^2 - a)(b - z^2)(z^2 + c)$. Here $0 < a < 1 < b < \sqrt{3}$ and $c \in (\sqrt{3}, 2)$.
\item For $C = 0$ one obtains Schuster's solution (\ref{schuster}). In this case $w(z) = -z^2 (z^4 - 3)$. Details of the derivation of (\ref{schuster}) can be found in~\cite{chandrasekhar}.
\item For $C \in (0,2)$ the polynomial $w(z)$ has 2 real roots. Its factorisation yields $w(z) = (z^2 + a)(z^2 + b)(c - z^2)$, where $a$, $b$, and $c$ are given as before (they also fall in the same ranges).
\item For $C = 2$ the polynomial $w(z)$ can be easily factored, i.e., $w(z) = - (z^2 - 2)(z^2 + 1)^2$. This case leads to Srivastava's solution (\ref{sri}).
\item For $C > 2$ the polynomial $w(z)$ has 2 real roots. In this case the factorisation is even simpler. We have $w(z) = (f - z^2)(z^4 + f z^2 + f^2 - 3)$, where
\begin{equation}
\label{cardano2}
f = A + 1/A, \;\;\; A = \left( \frac{1}{2} \left( C - \sqrt{C^2 - 4} \right) \right)^\frac{1}{3},
\end{equation}
but we will not need this form of $w(z)$.
\end{enumerate}
Out of the above 7 cases, only 3, namely (3), (5), and (7), require a separate discussion. A closer inspection (see Sec.~2.3) shows that solution (\ref{gh}) of Goenner and Havas is, in fact, valid in cases (5) and (7). In case (5) a formula based on Jacobian elliptic functions can be also provided.

\subsection{Solution for $-2 < C < 0$}

For $C \in (-2,0)$ Eq.~(\ref{dyn}) can be written as
\[ \frac{dz}{\sqrt{\left( z^2 - a \right) \left( b - z^2 \right) \left( z^2 + c \right)}} = \pm \frac{1}{2 \sqrt{3}}dt, \]
where $\sqrt{a} < |z| < \sqrt{b}$ (here $a$, $b$, and $c$ are given by  Eq.~(\ref{cardano})). Substituting $x = z/\sqrt{z^2 - a}$, $|x| \in (\sqrt{b}/\sqrt{b - a}, \infty)$, we obtain
\[ I = \int \frac{dz}{\sqrt{\left( z^2 - a \right) \left( b - z^2 \right) \left( z^2 + c \right)}} = - \int \frac{dx}{\sqrt{(b-a)x^2 - b} \sqrt{(a+c)x^2 - c}}. \]
It is convenient to introduce a rescaled variable $y = \sqrt{b - a} x / \sqrt{b}$, $ |y| \in (1, \infty)$. This yields
\[ I = - \frac{1}{\sqrt{(a+c)b}} \int \frac{dy}{\sqrt{y^2 - 1} \sqrt{y^2 - k^2}},  \]
where $k^2 = (b-a)c/(a+c)/b$. Noting that~\cite{abramovitz}
\[ \mathrm{arcdc}(x,k) = \int_1^x \frac{dy}{\sqrt{y^2 - 1}\sqrt{y^2 - k^2}}, \;\;\; 1 \leq x < \infty, \]
and returning to original variables, one finds a solution to Eq.~(\ref{le}) in the form
\begin{equation}
\label{case1}
\theta(\xi) = \pm \sqrt{\frac{a b y^2}{2 \xi \left( b y^2 - (b-a) \right)}}, \;\;\; y = \mathrm{dc} \left( \frac{1}{2} \sqrt{\frac{(a+c)b}{3}} \ln (B \xi), \sqrt{\frac{(b-a)c}{(a+c)b}} \right).
\end{equation}
The new integration constant $B$ corresponds to the scaling symmetry of Eq.~(\ref{le}), and $\mathrm{dc}$ is a subsidiary Jacobian elliptic function (in the standard Glaisher notation). Note that $|\mathrm{dc}(x,k)| \ge 1$ for $k \in [0,1]$, and thus solution (\ref{case1}) has no zeros.

\subsection{Solution for $0 < C < 2$}

For $C \in (0,2)$ Eq.~(\ref{dyn}) leads to the integral
\[ I = \int \frac{dz}{\sqrt{\left( z^2 + a \right) \left( z^2 + b \right) \left( c - z^2 \right)}}, \]
where $z \in (-\sqrt{c},\sqrt{c})$. It can be solved in a similar way. The substitution $y = \sqrt{(a+c)/a}(z/\sqrt{c-z^2})$, $y \in \mathbb R$ yields
\[ I = \frac{1}{\sqrt{(a+c)b}} \int \frac{dy}{\sqrt{1+y^2} \sqrt{1 + \left(1 - k^2 \right) y^2}}, \]
where again $k^2 = (b-a)c/(a+c)/b$. The above integral gives rise to another elliptic function, namely~\cite{abramovitz}
\[ \mathrm{arcsc}(x,k) = \int_0^x \frac{dy}{\sqrt{1 + y^2}\sqrt{1 + \left(1 - k^2 \right)y^2}}, \;\;\; -\infty < x < \infty. \]
The solution can be written in the form
\begin{equation}
\label{case2}
\theta(\xi) = \pm \sqrt{\frac{a c y^2}{2 \xi \left( a y^2 + a + c \right)}}, \;\;\; y = \mathrm{sc} \left( \frac{1}{2} \sqrt{\frac{(a+c)b}{3}} \ln (B \xi), \sqrt{\frac{(b-a)c}{(a+c)b}} \right).
\end{equation}
Here again $a$, $b$, and $c$ are given by (\ref{cardano}), $B$ is an integration constant, and $\mathrm{sc}$ denotes another subsidiary Jacobian elliptic function. Note that the expression for $\theta(\xi)$ is regular, even for $y \to \pm \infty$. The sign in (\ref{case2}) has to be changed as the solution curve passes through a zero.

\subsection{Solution for $C > 2$}

For $C > 2$ and $f$ given by Eq.~(\ref{cardano2}), the integral
\[ I = \int \frac{dz}{\sqrt{(f - z^2)(z^4 + f z^2 + f^2 - 3)}}, \]
with $z \in (-\sqrt{f}, \sqrt{f})$ can be computed in a manner similar to that used in preceding sections, that is by removing the term $(f - z^2)^{-1/2}$ from the integrand. However, it turns out that the substitution $z = \pm \sqrt{C/(s - 1)}$, $ s \in [1+C/f, \infty)$  in the expression
\[ I = \int \frac{dz}{\sqrt{-z^6 + 3z^2 + C}} \]
yields an even simpler result. One obtains
\[ I = \mp \int \frac{ds}{\sqrt{4 s^3 - 12 s - 4(C^2 - 2)}}. \]
Equation (\ref{dyn}) can can be now integrated by noticing the following identity for the Weierstrass elliptic function $\wp (z; g_s, g_3)$ with invariants $g_2$ and $g_3$ \cite{abramovitz}:
\[ z = \int_{\wp (z; g_s, g_3)}^\infty \frac{ds}{\sqrt{4 s^3 - g_2 s - g_3}}. \]
Returning to the original variables, we can write the solution in the form
\begin{equation}
\label{case3}
\theta(\xi) = \pm \frac{\sqrt{C}}{\sqrt{2\xi}\sqrt{\wp\left( \ln(B \xi)/(2\sqrt{3}); 12, 4(C^2 - 2) \right) - 1}},
\end{equation}
where $B$ is an integration constant. This result is actually valid for all $C > 0$ (it includes solution (\ref{case2})), and it is equivalent to solution (\ref{gh}) obtained by Goenner and Havas in~\cite{goenner}. The latter follows straightforwardly from the identity
\[ \wp \left(\lambda z; \lambda^{-4}g_2, \lambda^{-6} g_3 \right) = \lambda^{-2} \wp(z; g_2, g_3), \]
which holds for any constant $\lambda \neq 0$ \cite{abramovitz}.

\begin{figure}[t]
\begin{center}
\includegraphics[width=\textwidth]{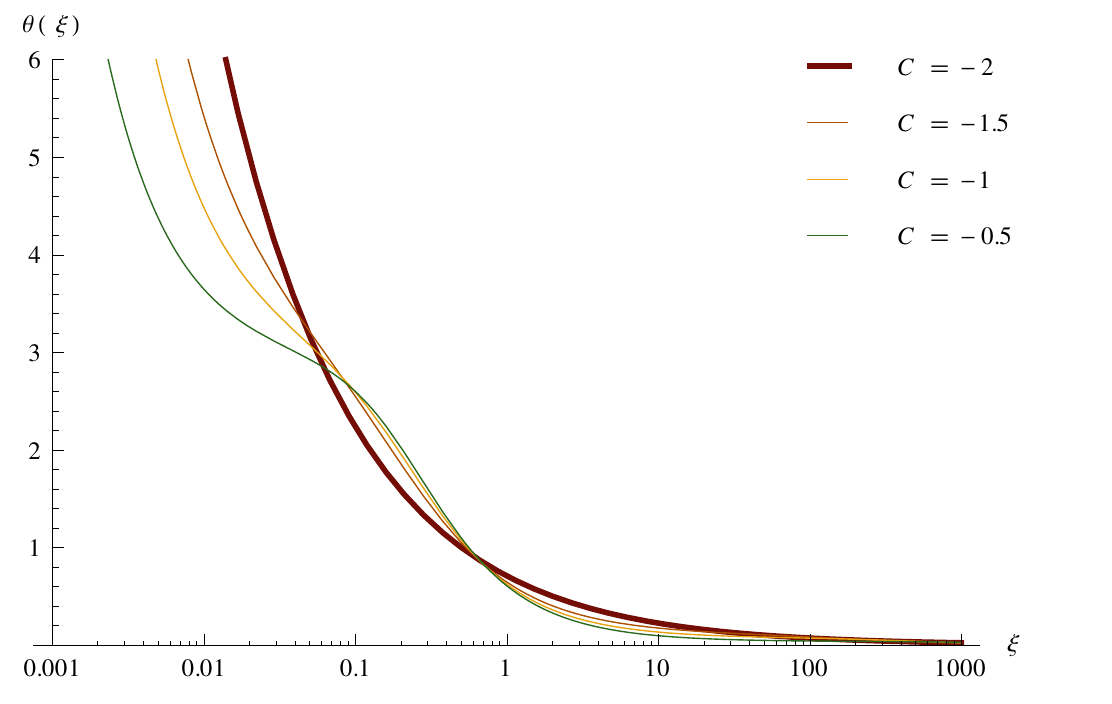}
\end{center}
\caption{\label{c_negative} Plot of solutions $\theta(\xi)$ for $C = -2$ (thick line) and $C = -1.5, -1, -0.5$. Only positive solutions are presented. The radial variable $\xi$ is plotted in the logarithmic scale.}
\end{figure}

\begin{figure}[t]
\begin{center}
\includegraphics[width=\textwidth]{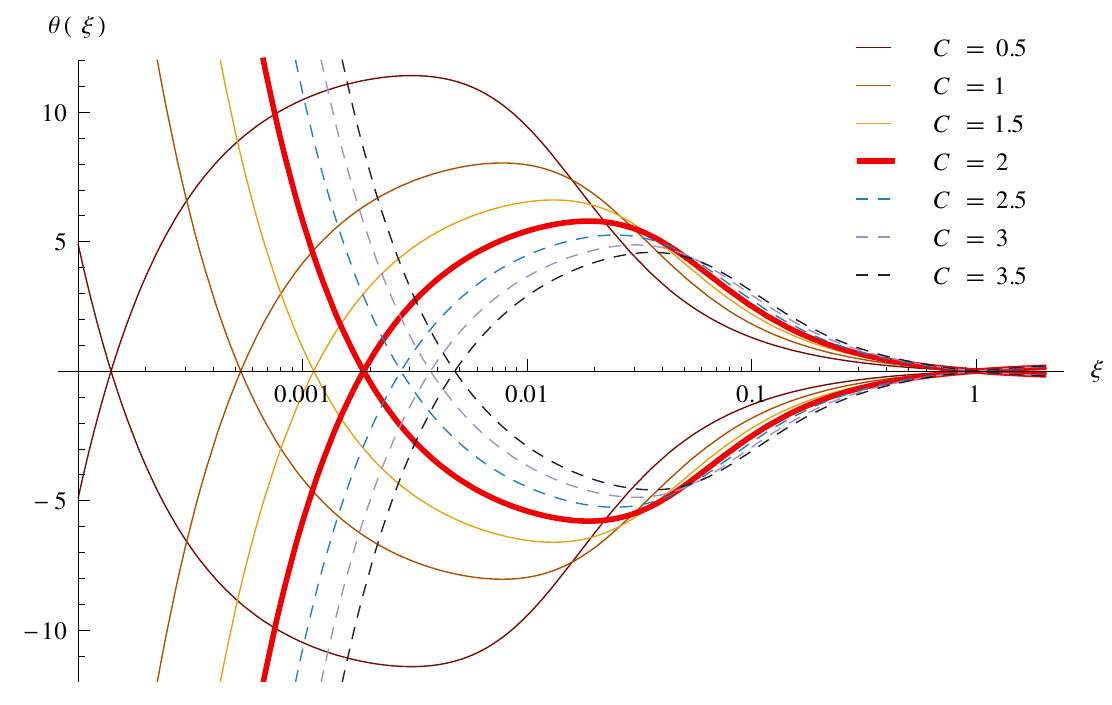}
\end{center}
\caption{\label{c_positive} Plot of solutions $\theta(\xi)$ for $C = 0.5, 1, 1.5$ (solid lines), $C = 2$ (thick line), $C = 2.5, 3, 3.5$ (dashed lines). The radial variable $\xi$ is plotted in the logarithmic scale.}
\end{figure}

\section{Discussion}

Solutions of Eq.~(\ref{le}), or equivalently Eq.~(\ref{dyn}), belong to three, essentially different types. Solutions for $C \in [-2,0)$ are singular at $\xi = 0$, and have no zeros. Schuster's integral, that is the solution for $C = 0$, is regular everywhere, and its sign is also definite. Solutions for $C > 0$ are all oscillating, and the frequency of the oscillations grows infinitely as $\xi \to 0$. This can be easily demonstrated for the Srivastava's solution (\ref{sri})---it has zeros at $\xi = \exp(2 k \pi)$, $k \in \mathbb Z$.

More interestingly, all solutions for $C \in (-2,0)$ and $C > 0$ have a non-trivial, discrete scaling symmetry, which follows from the periodicity of elliptic functions. We will demonstrate this behaviour for solutions (\ref{case1}). Applying the scaling transformation $S$ to (\ref{case1}) one obtains
\[ \theta(\xi/\lambda)/\sqrt{\lambda} = \pm \sqrt{\frac{a b y_\lambda^2}{2 \xi \left( b y_\lambda^2 - (b-a) \right)}},\]
where
\[ y_\lambda = \mathrm{dc} \left( \frac{1}{2} \sqrt{\frac{(a+c)b}{3}} \ln (B \xi/\lambda), \sqrt{\frac{(b-a)c}{(a+c)b}} \right). \]
The function $\mathrm{dc}(x,k)$ is periodic, with the period equal to $2 K(k)$, where $K(k)$ is the complete elliptic integral of the first kind. Thus, for
\begin{equation}
\label{lambda}
\lambda = \exp \left( \frac{4 \sqrt{3} m}{\sqrt{(a+c)b}} K \left( \sqrt{\frac{(b-a)c}{(a+c)b}} \right) \right), \;\;\; m \in \mathbb Z,
\end{equation}
we have $y_\lambda = y_1$, and $\theta(\xi/\lambda)/\sqrt{\lambda} = \theta(\xi)$. Clearly, solutions (\ref{case1}) are fixed points of $S$ for the values of $\lambda$ given by (\ref{lambda}). The same formula holds for solutions (\ref{case2}), and an analogous one can be obtained for Eq.~(\ref{case3}).

Sample solutions for $C < 0$ and $C>0$ are plotted in Figs.~\ref{c_negative} and \ref{c_positive}, respectively. Note that the $\xi$ axis is logarithmic on both graphs. Figure \ref{c_negative} shows the scale invariant solution (\ref{fixed})---it is depicted with a thick line, and solutions given by Eq.~(\ref{case1}) for $C = -3/2$, $-1$, and $-1/2$. The integration constant $B$ appearing in (\ref{case1}) is chosen in the way assuring that all solutions intersect at $\xi = 1/2$. Figure \ref{c_positive} shows the solution of Srivastava (thick line), solutions given by Eq.~(\ref{case2}) for $C = 1/2$, 1, and 3/2 (thin solid lines), and solutions given by Eq.~(\ref{case3}) for $C = 5/2$, 3, and $7/2$ (dashed lines). Here the constant $B$ is set to $B = 1$.

Newly obtained solutions can be used in composite stellar models on the same footing as Srivastava's solution~\cite{murphy83}. Lane--Emden equations appear also in the description of the steady state of some field-theoretical models. In this respect, a good example is provided by works dealing with semilinear wave equations with focusing power-law nonlinearities---see e.g.~\cite{bizon1, bizon2, kycia}.

Many variants and generalisations of Lane--Emden equations appear in different areas of physics (a large collection of relevant references is provided in \cite{goenner}). These generalised versions were studied by many authors, and analytic solutions have been obtained as well. Recent literature on this subject includes also \cite{khalique, muatjetjeja, gorder}.

\section*{Acknowledgements}

I would like to thank Piotr Bizo\'{n}, Edward Malec, Andrzej Odrzywo\l{}ek, Pawe\l{} Biernat, Rados\l{}aw Kycia and Maciej Maliborski for discussions and invaluable help in improving this article.

\end{document}